\documentclass[12pt]{article}

\usepackage[margin=2.5cm]{geometry}

\usepackage{amsmath,amsthm,amsfonts,amscd,amssymb,bbm,mathrsfs,enumerate,url}
\usepackage{csquotes}
\usepackage{comment}

\usepackage{graphicx,tikz}
\usepackage[pdfborder={0 0 0}]{hyperref}
\usetikzlibrary{matrix,arrows}
\usetikzlibrary{decorations.pathreplacing}
\usetikzlibrary{decorations.pathmorphing}
\usetikzlibrary{patterns,fadings}

\def\RR{{\mathbb R}}

\def\1{{\mathbbm 1}}

\def\diff{{\rm Diff}}

\def\diffs1{\diff(S^1)}

\def\psl2r{{\rm PSL}(2,\RR)}
\def\sl2r{{\rm SL}(2,\RR)}
\def\su11{{\rm SU}(1,1)}
\def\2dmob{{\overline{\psl2r}\times\overline{\psl2r}}}
\def\<{\langle}
\def\>{\rangle}

\theoremstyle{remark}

\newcommand{\svh}{\mathrm{Sv/h}}

\title{Comments on
``Individual external dose monitoring of all citizens of {D}ate {C}ity by
  passive dosimeter 5 to 51 months after the {F}ukushima {NPP} accident (series):
  {II}. {P}rediction of lifetime additional effective dose and evaluating the
  effect of decontamination on individual dose.''
}
\date{} 
\author{
{\bf Yoh Tanimoto}
\\
   Dipartimento di Matematica, Universit\`a di Roma ``Tor Vergata''\\
   email: {\tt hoyt@mat.uniroma2.it} \\
{\bf Yutaka Hamaoka} \\
Faculty of Business and Commerce, Keio University\\
email: {\tt hamaoka@fbc.keio.ac.jp}\\
{\bf Kyo Kageura} \\
Graduate School of Education, University of Tokyo\\
email: {\tt kyo@p.u-tokyo.ac.jp}\\
{\bf Shin-ichi Kurokawa} \\
The High Energy Accelerator Research Organization (KEK), Tsukuba \\
email: {\tt shin-ichi.kurokawa@kek.jp}\\
{\bf Jun Makino} \\
Department of Planetology, Graduate School of
  Science, Kobe University\\
email: {\tt makino@mail.jmlab.jp}\\
{\bf Masaki Oshikawa} \\
Institute for Solid State Physics, University of Tokyo\\
email: {\tt oshikawa@issp.u-tokyo.ac.jp}\\
}
\begin{document}

\begin{center}
{\LARGE Comments on the second paper of the series ``Individual external
dose monitoring of all citizens of Date City by passive dosimeter
5 to 51 months after the Fukushima NPP accident''}
\end{center}

\vspace{2cm}
\noindent
\textbf{Shin-ichi Kurokawa}

\noindent
High Energy Accelerator Research Organization,KEK

\vspace{0.5cm}
\noindent
Dear Sir

\vspace{0.5cm}
\noindent
Makoto Miyazaki and Ryugo Hayano have published two papers of the
series, ``Individual external dose monitoring of all citizens of Date
City by passive dosimeter 5 to 51 months after the Fukushima NPP accident
(series)''.\cite{key-9}\cite{key-10} I point out serious inconsistencies
in the second paper. To make the argument simpler, I only discuss
the case for zone A.

In order to evaluate the lifetime additional doses the authors determine
the ambient dose reduction function $f(t)$, which is supposed to
be normalized to unity at $t=0.65y$, multiply it by the mean ambient
dose rate $\dot{H}_{10}^{*A}(0.65)$ and the coefficient $c^{A}$
( the average of individual doses divided by ambient doses), and integrate
the product over time.

I firstly point out that the formula (1) of the paper that defines
$f(t)$ is incorrect. The formula does not give $f(0.65)=1$. A normalization
factor is missing.

The authors state that $\dot{H}_{10}^{*A}(0.65)$$=2.1$ $\mu{\rm Svh}^{-1}$
and $c^{A}=0.10$, which contradict figure 6, where the curve $\dot{H}_{p}^{A}(t)$
is apparently drawn by using the product of $\dot{H}_{10}^{*A}(0.65)$
and $c^{A}$ as 0.33 $\mu{\rm Svh}^{-1}$.

In the caption of figure 7, the authors write that ``glass-badge data
and the reduction curve shown in figure 6 are converted to cumulative
dose distributions and $H_{p}^{A}(t)$, respectively.'' In figure
6 the curve $\dot{H}_{p}^{A}(t)$ is lower than or almost equal to
the median values of the individual dose rate distributions, while
in figure 7 the curve $H_{p}^{A}(t)$ is higher by about 15\% than
the median values of the cumulative dose distributions. The authors
do not show any reason why this happens. 

At the time of $t=3y$ (36 months after the accident), the cumulative
dose shown by the curve $H_{p}^{A}(t)$ in figure 7 is 2.7 mSv and
that of median value is 2.4 mSv. I integrated the area below the curve
in figure 6 by using the properly normalized reduction function and
$\dot{H}_{p}^{A}(0.65)=0.33$ $\mu{\rm Svh}^{-1}$ from $t=0.39y$ to $t=3y$ and obtained the
value of 4.8 mSv. These three values must be equal or very similar
to each other. The discrepancies are very large and the reason of
this discrepancies are not shown in the paper.

The increase of the median of cumulative doses from 29 month to 38
month in figure 7 is measured to be between 0.18 and 0.12 mSv/3month. Since the reading
of glass-badge is digitized by 0.1 mSv step, the increase of dose
per three months must be an integer multiple of 0.1 mSv, when the number of the participants is odd, such as 425 in this case.  The increase
of 0.18-0.12 mSv/3month violates this principle. Let me point out that
in figure 6 median doses per 3 months in this period are 0.3 mSv,
which is twice as large as 0.15 mSv/3month.

The authors write that ``By extrapolating $H_{p}^{i}(t)$ to $t=70y$
(and by adding the estimated external dose for the initial 4 months)
we estimated the additional lifetime dose for each zone.'' They obtained
18 mSv for zone A. I also calculated the median lifetime dose for
zone A by using the $\dot{H}_{p}^{A}(0.65)=0.33$ $\mu{\rm Svh}^{-1}$
and have found that it is not 18 mSv but 26 mSv. 

In order to get 18 mSv for the lifetime dose, $\dot{H}_{p}^{A}(0.65)$
must be 0.22 $\mu{\rm Svh}^{-1}$. Moreover, in order to get the curve
and the median value of the cumulative doses of figure 7, $\dot{H}_{p}^{A}(0.65)$
must be 0.19 $\mu{\rm Svh}^{-1}$ and $0.17$ $\mu{\rm Svh}^{-1}$,
respectively. These three values differ significantly from $0.33$
$\mu{\rm Svh}^{-1}$.

I add three more comments : (i) the phrase ``mean ambient dose $\dot{H}_{10}^{*A}(0.65)$''
(under equation (2)) is inconsistent with the phrase ``median grid
dose $\dot{H}_{10}^{*A}(0.65)$'' (figure 6 caption, where it is wrongly
written as $\dot{H}_{10}^{iA}(0.65)$ ), (ii) according to equation
(2), the curve in figure 7 should not be $H_{p}^{A}(t)$ but $H_{p}^{A}(t)-I$,
and (iii) if $I$ is the ``dose during the first 4 months'', then
the lower limit of integration in equation (2) should be $t=0.33y$
rather than ``$t=0.39y$'' (= 4.7 months).

In summary, I have found serious inconsistencies in the paper, which
prevent me from getting reliable information from the paper.

I am grateful to Professor Shin Takagi for his valuable discussion and comments to the point.

Yours sincerely,

\vspace{0.5cm}
\noindent
\textbf{Shin-ichi Kurokawa}

\noindent
High Energy Accelerator Research Organization, KEK

\maketitle
% \begin{abstract}
% 
% \end{abstract}

\noindent
Dear Sir,

% One of us (SK) has submitted a letter \cite{Kurokawa18} on the paper by M.\! Miyazaki and R.\! Hayano
% published by JRP \cite{MH17}, pointing out various inconsistencies therein.

In this Letter, we point out inconsistencies, obvious mistakes and inappropriate statements
in \cite{MH17}, which were not discussed in the earlier Letter~\cite{Kurokawa18} by
one of us (SK).
Throughout this Letter, ``Fig. X'' refers to a figure in Ref.\! \cite{MH17} unless otherwise specified. 

\begin{enumerate}[{(}1{)}]
 %\item It is claimed that ``\textit{[T]he glass badge data and the GIS information of the subjects
 %were anonymized by Date City, and was provided to the authors}''.
 %This appears inconsistent with the statement in the other paper of the series \cite{MH16}
 %that ``\textit{[T]he geocoded household addresses of the glass-badge monitoring
 %participants were pseudo-anonymized by rounding both longitude and latitude coordinates to $1/100$
 %degrees prior to data analyses}'', which means that it was %the authors who did pseudo-anonymization (note that the %glass-badge data are the same).
 
 \item Section 1 gives a summary of the findings in the first paper of the series \cite{MH16}, yet
 the statement ``\textit{[T]he obtained coefficient of $0.15$ did not change with time, across the six airborne surveys
 conducted between November 2011 (4th survey) and November 2014 (9th survey)}''
 is not supported by either a statement or an appropriate statistical analysis of the data for each airborne survey. Specifically, the authors did not present values of the coefficient $c$ for each airborne survey, but only an aggregated arithmetic mean over all surveys \cite{MH16}.
 
 \item Fig.\! 1 contains several anomalies. 
 According to our examination, the original airborne monitoring data \cite{MEXT_airborne} contains $4{,}150$ grid points in Date city\footnote{
The area of Date City is $265\,\mathrm{km}^2$ and airborne does is estimated for each $250\,\mathrm{m}$ grid points or $4 \times 4$ grid per $1\,\mathrm{km}^2$, thus we expect around $265 \times 16 = 4{,}240$ data points. Although, for this letter we utilized original data in Japanese\cite{MEXT_airborne}, English version data is available at the following site. \url{https://emdb.jaea.go.jp/emdb/en/portals/b1010301/}

For example, ``Air Dose Rate Results of the Fourth Airborne Monitoring Survey ( Decay correction: November 5, 2011 )'' for Fukushima is obtained from the following link in the unified format. In the file, $4{,}150$ records are listed for ``Date city.''
\url{https://emdb.jaea.go.jp/emdb/assets/site_data/en/csv_utf8_unifiedformat/1010301004/1010301004_07.csv.zip} }.
 We made a figure using all $4,150$ points (the top panel of Fig \! \ref{fig:MH2Fig1} of this Letter).
 \begin{itemize}
 \item In spite of the fact that the range of the ambient dose for the 4th monitoring\footnote{
  Using above mentioned file, we calculated descriptive statistics and subtracted back ground dose of $0.04\,\mu\svh$, then obtained range of $(0.39, 6.76)\,\mu\svh$. For the later survey, the same analysis was conducted with the archived data.} after subtracting background of $0.04\,\mu\svh$ is $(0.39, 6.76)\,\mu\svh$
  (see the top panel of Figure 1 of this paper, where the $x$-coordinates of the plots range from $0.39$ to $6.76$), many points under $0.39\,\mu\svh$ are plotted and points over 2.7$\,\mu\svh$ are missing in the Fig.\! 1. The same is observed for the vertical axis; for example, the range of airborne dose for the 5th monitoring is $(0.24, 3.66)\,\mu\svh$, but many points below $0.24 \,\mu\svh$ are plotted and points larger than $2.5\,\mu\svh$ are not shown. 

 \item In the original airborne monitoring data, the points higher than $1.0\,\mu\svh$ are discretized into steps of $0.1\,\mu\svh$.  However, such a discretisation is not present in Figure 1 of \cite{MH17}. This is evident by comparing the top panel of Fig.\! \ref{fig:MH2Fig1} of this Letter to the original plot shown as the bottom panel.
  
  \item In Fig.\! 4a of  \cite{MH16}, the bins whose ambient dose rate is smaller than $0.35\,\mu\svh$ do not have any subjects.  This is inconsistent with Fig.\! 1 where there are points under $0.39\,\mu\svh$.
  \end{itemize}
%This paper \cite{MH17} claims that airborne monitoring dose can be utilized to predict individual dose, but the treatment of the fundamental data is unclear and unreliable.

\begin{figure}[ht]
\centering
\includegraphics[width=10cm]{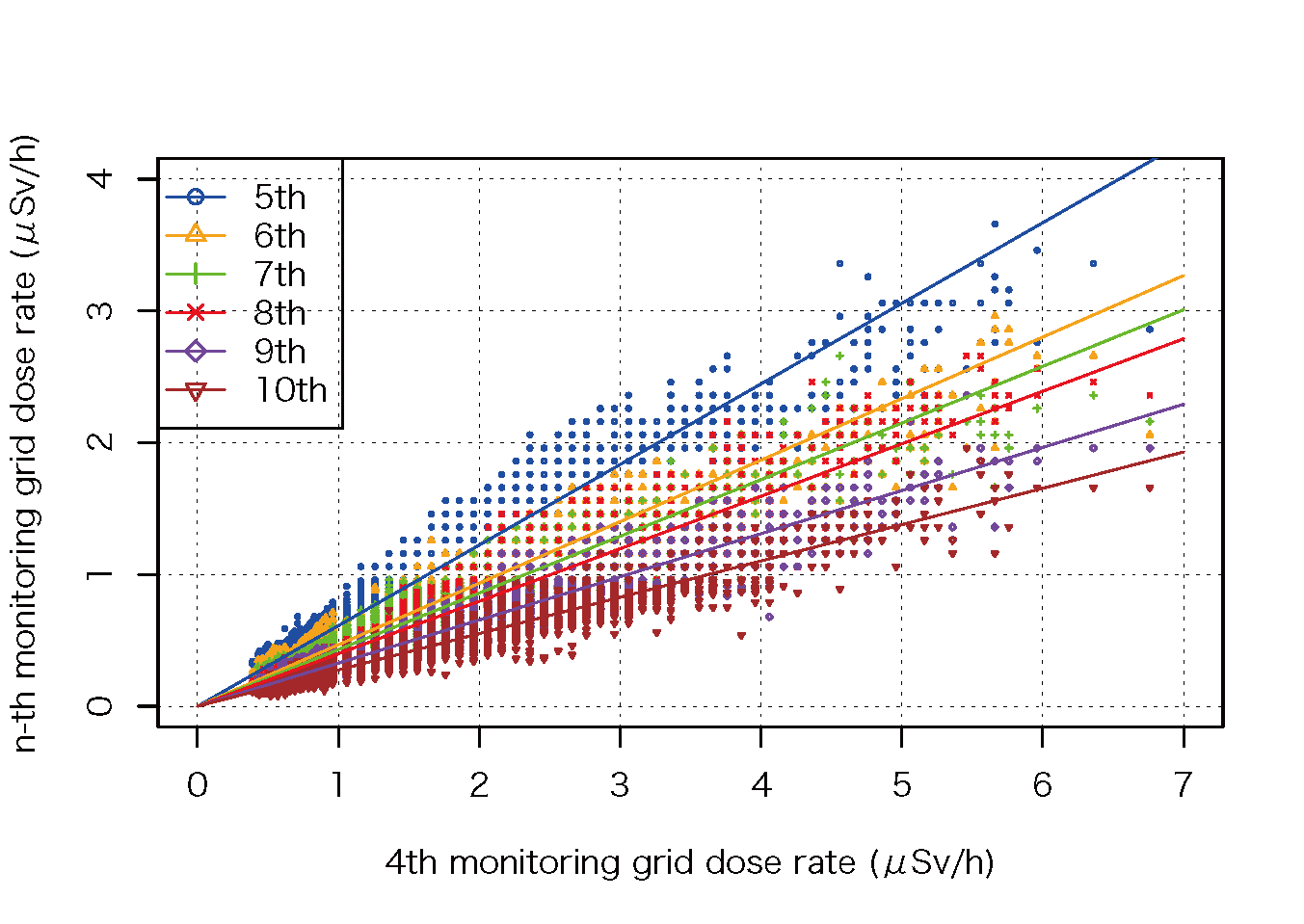}\\
\includegraphics[width=10cm]{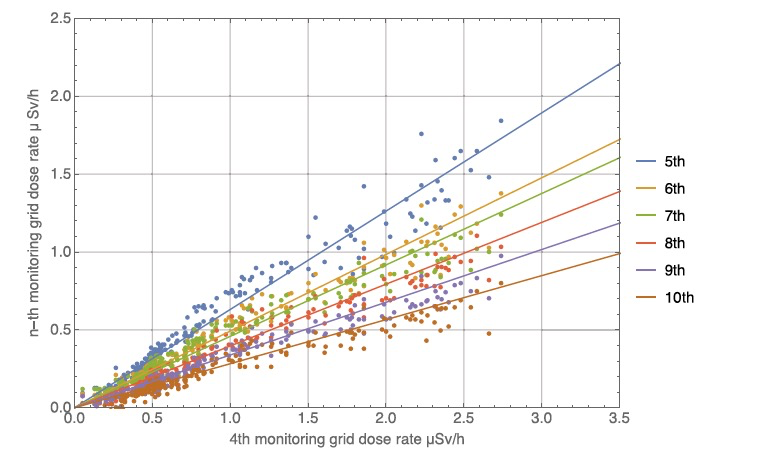}
 \caption{Comparison between Reproduced Scatter Plot with full-data and Fig.1 in \cite{MH17}. The top panel was plotted using original data \cite{MEXT_airborne} with regression lines ($n$-th air monitoring dose dose was regressed on the 4th monitoring dose without intercept). The bottom panel was taken from \cite{MH17} for comparison.} 
\label{fig:MH2Fig1}
\end{figure}

 %\item Fig.\! 1 contains several anomalies.
% \begin{itemize}
%     \item Each survey in the original data \cite{MEXT_airborne} contains around $4{,}150$ points, but the shown plots clearly do not.
%     \item The original data contain points higher than $2.7\,\mu\svh$, but they are not shown.
%     \item The original data of the 4th survey do not contain points less than $0.39\,\mu\svh$ (after subtracting the background radiation $0.04\,\mu\svh$, but the shown plots do.
%     \item The points higher than $1.0\,\mu\svh$ in the original data are discretized into steps of $0.1\,\mu\svh$, but the the shown plots are not. 
%  \end{itemize}
% The authors do not explain how they plotted the data.

 \item It is claimed in Section 2.2, without showing data, that $c^{\mathrm A} = 0.10$
 and $\dot H_{10}^{*A}(0.65) = 2.1\,\mu \svh$, whose
 product is $c^{\mathrm A} \cdot \dot H_{10}^{*A}(0.65) = 0.21\,\mu\svh$. We strongly suspect that this is false:
 \begin{itemize}
 \item The households in Zone A are shown in Fig.\! \ref{fig:MH2Fig3colored} of this Letter. Colored points corresponds to the households in Fig.\! 3a of \cite{MH16}, and colors show the ambient dose rate at time $t=0.65\,\mathrm{y}$.
 The ambient dose rates of most of these households were
 higher than $2.5\,\mu\svh$, therefore, it is unlikely that
 the average $\dot H_{10}^{*A}(0.65)$ was $2.1\,\mu\svh$, but it should be somewhere between $2.5\,\mu\svh$ and $3.5\,\mu\svh$.
 \begin{figure}[ht]
 \centering
\includegraphics[width=10cm]{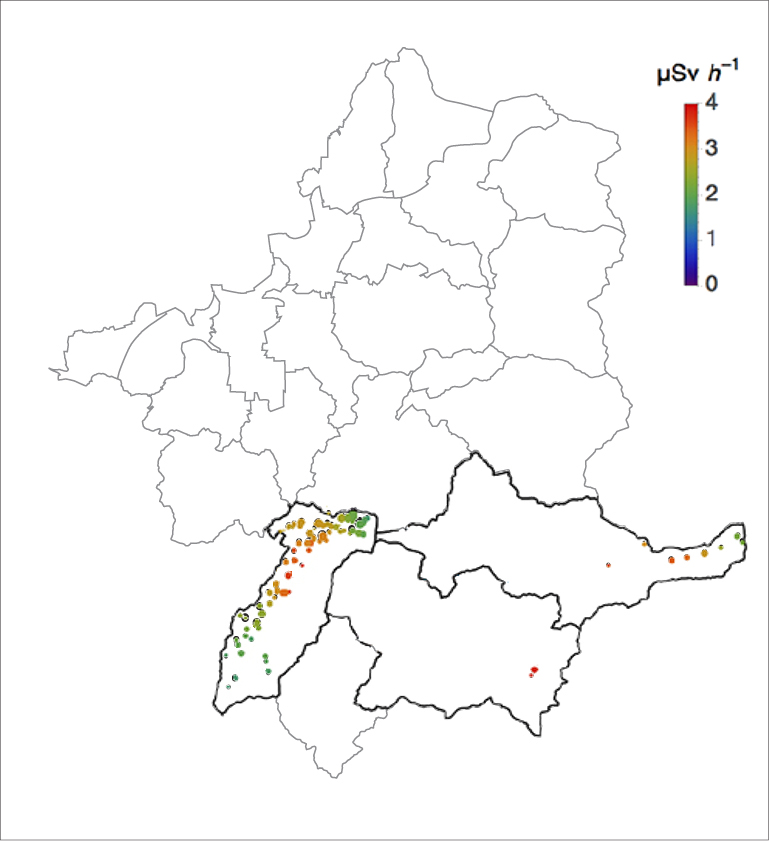}<
 \caption{The points corresponding to the households in Fig.\! 3 are overlaid by the colored points in Fig.\! 3a of \cite{MH16}, where ``the area of each circle is drawn proportionally to the number of participants within each the grid cell''. At least half of them are in the grid cells with $2.5\,\svh$ or higher.} 
\label{fig:MH2Fig3colored}
\end{figure}

 \item If we assume the log-normal distribution of \cite[Fig.\! 5]{MH16} at each period, the mean value must be around $1.15$ times higher than the median. For the curve in Fig.\! 6 to pass the mean value at $7$-th month, the product $c^{\mathrm A} \cdot \dot H_{10}^{*A}(0.65)$ would have to be about $0.42\,\mu\,\svh$ (see also \cite[P.14]{Hayano151015}, where a plot very similar to Fig.\! 6 is shown with additional orange dots. We suspect that these dots are the average dose rate in each period, and its value at 2011 Q3 (the 7th month) is about $0.95\,\mathrm{mSv}/3\mathrm{month} = 0.43\,\mu\svh$). This suggests that the claimed value $c^{\mathrm A} = 0.10$ is considerably underestimated. By looking at Fig.\! 4a of \cite{MH16}, if we pick up the part where the ambient dose rate is higher than $2.5\,\mu\svh$ and hence should contain most of the households in Zone A, it is reasonable to assume that $c^{\mathrm{A}} \sim 0.15$.
 \end{itemize}
 The product $c^{\mathrm A}\cdot \dot H_{10}^{*A}(0.65)$ appears as a factor, therefore,
 any error in it leads to an error in the estimated lifetime doses in Zone A, which is a main result of the paper.

 \item All three panels of Figs.\! 5 have box-and-whisker plots for the $5$-th month (August 2011).  These plots should not be included in the figures, since it is written in the paper ``\textit{we selected the subjects who held the glass badges continuously from 2011Q3 to 2015Q1 (from September 2011 to June 2015)}'' (in Section~2.2).  
  
 Moreover, the statement in Section 4 ``\textit{Date City, which began measuring personal doses for schoolchildren and residents living in relatively high-dose areas from August 2011}'' is imprecise: children in the survey in August 2011 were those aged 15 or younger, and there were about $100$ pregnant women as well, but most of the adult residents of the area started to participate in September 2011, see \cite[Table 1]{MH16}.
 
 \item In Fig.\! 5c, the lower whisker decreases between $14$-th and $17$-th months,
 $23$-rd and $26$-th months, and $44$-th and $47$-th months.
 This must be impossible because the figure is supposed to show the cumulative doses  of those ``\textit{who continuously held glass badges during the study period}'', i.e.\! of the same participants. This suggests a serious error in handling the data, or an unusual way of plotting. 
 
 \item The authors conclude, by picking only two periods among 10 (before and after the decontamination)
 and assuming a single reduction function throughout the whole period,
 that ``\textit{effects of decontamination on the reduction of individual doses were not evident}''.
 This conclusion is unreasonable: the authors should have compared the dose rates before and
 after the decontamination and fitted each period by an {\it a priori} different function.
 Indeed, one of the authors (R. Hayano) reported in a symposium on September 13, 2015 that there were effects of decontamination in Zone A by showing the same graph \cite[13:00$\sim$]{Hayano1509Dialogue} and they also wrote that the effect of the decontamination in Zone A was $\sim 60\%$ in  \cite[P.14]{Hayano151015}.

 \item In Figs.\! 5 and 6, a majority of the participants in Zone A remained evacuated in 2014 \cite[P.97, Table]{Date3Years}.
 This means that
 the ambient dose rates associated with their addresses are not correlated with the actual dose
 rates they receive, and it makes little sense to claim based on such data that effects of decontamination were not evident.
 Furthermore, most participants in Zones B and C are children younger than or equal to 15 years old (see \cite[Table 1]{MH16}),
 and their parents often tend not to let them stay outside for a long time. Analyses without considering
 these factors distort the estimated lifetime doses.

 \item In Fig.\! 7, the number of participants is $425$, while there are more than or nearly $10$ outliers
 above each upper whisker. This is impossible if the upper whisker represents the $99$-th percentile.
 Rather, the upper whiskers are suspected to correspond to the $90$-th percentile by comparison with \cite[Fig.\! 5]{MH16} where the $90$-th percentile is about $2.1$ times of the median. This affects the claimed $99$-th percentile
 of the estimated lifetime doses.
 This argument applies also to Fig.\! 5, Zone A with $476$ participants.
 
 In addition, if we subtract the initial dose of $1.4\,\mathrm{mSv}$, three curves in each figure of Figs.\! 5, high, middle, and low curves, have a special relation to each other; the ratio between high/middle is about 2, and also the ratio between
 middle/low is about 2. From \cite[Fig.\! 5]{MH16}, we know that the $90$-th percentile is about twice as large as the median ($50$-th percentile) and $10$-th percentile is about half of the median.  We suspect that the high curve corresponds to $90$-th percentile, the middle to the median and the low to the $10$-th percentile. The fact that high curves pass close to the upper end of the whiskers and low curves do not pass the lower end of the whiskers and go higher corroborates our suspicion that the upper end of whiskers are not 99-th percentile but 90-th percentile and that low curve corresponds to $10$-th percentile, and not to $1$-st percentile as described in the caption of Fig.\! 5.
 
 \item The accumulated doses in Fig.7 should be obtained directly summing the doses of three months shown as $\mathrm{mSv}$ in the database, or by summing the dose rate shown $\mu\svh$ in Fig.\! 6 multiplied by $24$(hour) $\times$ $91$(day); however, as we pointed out in \cite{Kurokawa18}, the doses in Fig.\! 7 do not coincide with the doses calculated as described above and amount to only 46\% \cite{KT19-1}.
 Furthermore, there is a discrepancy between the integrated curve of $\dot H_{10}^{*A}$ in Fig.\! 6 (note that the curve in Fig.\! 6 is drawn using the value $c^{\mathrm A} \cdot \dot H_{10}^{*A}(0.65) = 0.33\,\mu\svh$, instead of the claimed values $\dot H_{10}^{*A}(0.65) = 2.1\,\mu\,\svh$ and $c^{\mathrm A} = 0.1$) and the cumulative doses $H_{10}^{*A}$ in Fig.\! 7: more precisely, the discrepancy is a factor of $0.58$.
 These two shrinkage factors, $0.46$ and $0.58$, cannot be explained by a single mistake (e.g. we pointed out that the factor of $0.46$ could be due to mistaken unit $\mu\svh$ instead of $\mathrm{mSv}/3\mathrm{months}$ \cite{KT19-1}, but this could not explain the factor of $0.58$).
 Moreover, the glass-badge data are discretized into steps of $0.1\,\mathrm{mSv}/3\mathrm{month}$ while the minimum step in Fig. 7 \! is $0.1 \mathrm{mSv}/3\mathrm{month}$ $\times$  $0.46$ (see Fig. 3 of this letter), therefore, the plots of Fig.\! 7 must be wrong.

 \item
  \begin{figure}[ht]
 \centering
 \includegraphics[width=16cm]{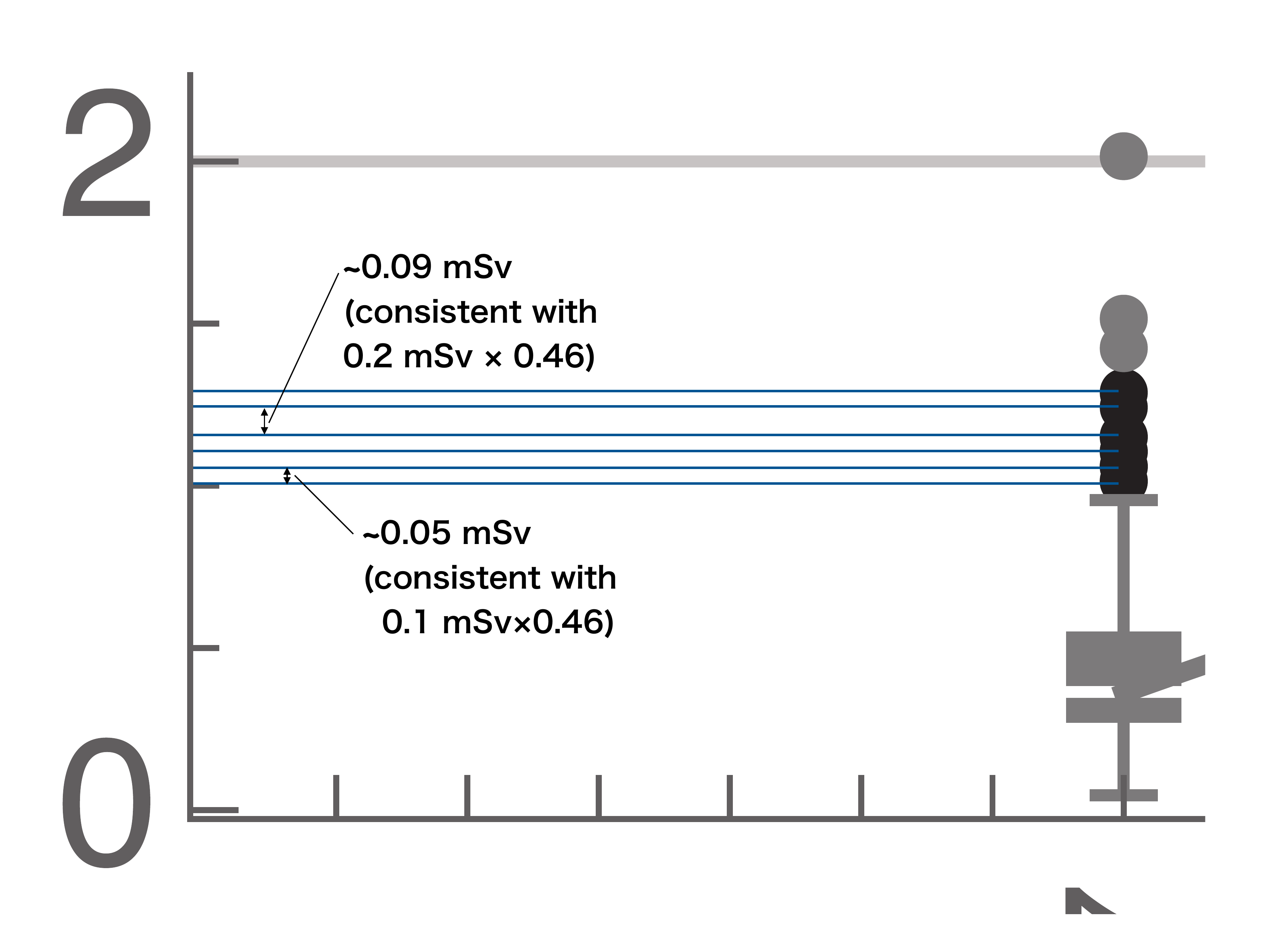}
    \caption{The minimal steps in Fig.\! 7 are $0.1\,\mathrm{mSv}$ $\times$ $0.46$, while the minimal increase of the doses registered with a glass-badge should be $0.1\,\mathrm{mSv}$. }
  \label{fig:7steps}
\end{figure}

 Note also that the participants of Fig.\! 5a are the residents in Zone A who held the glass-badge continuously in the period 2011Q3 to 2015Q1, while those of Fig.\! 7 held the glass-badge in the period 2011Q3 to 2014Q1 and their houses were subject of decontamination in 2012 Q3. It is natural to expect that most participants are common in these figures. If so, Fig.\! 5 has the same problem: compared with Fig.\! 6, individual doses are smaller by a factor of $0.55$ and the integrated curve is smaller by a factor of $0.70$ \cite{KT19-1}.
 \label{item.conversion}

 \item Let us assume that the glass-badge data of Fig.\! 5 are shrunken by a factor of $0.55$, as we argued above.
 On the other hand, the outliers are discretized into steps of about $0.1\,\mathrm{mSv}$ (see Fig.\! \ref{fig:mh2fig5steps} of this Letter). This means that the original data would have to be discretized into steps of about $0.2\,\mathrm{mSv}$, which is very unlikely because the minimal increase of a glass-badge is $0.1\,\mathrm{mSv}$. This suggests an error in plotting in Fig.\! 5.

  \begin{figure}[ht]
 \centering
 \includegraphics[width=15cm]{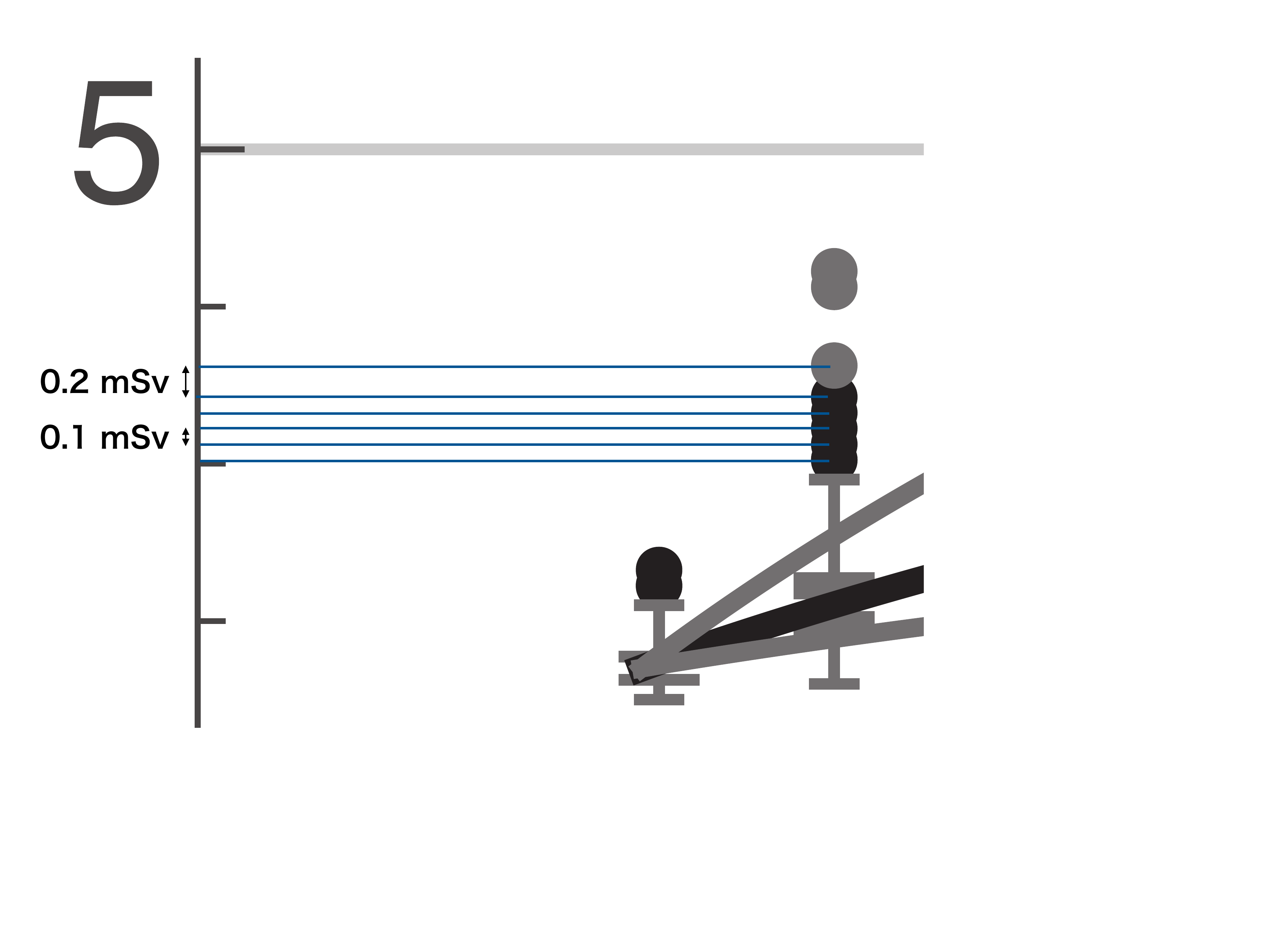}
    \caption{A part of Fig.\! 5a, enlarged. The minimal distance between outliers is about $0.1\,\mathrm{mSv}$. (The scale on the vertical axis is $1\,\mathrm{mSv}$)}
  \label{fig:mh2fig5steps}
\end{figure}

 \item
  \begin{figure}[ht]
 \centering
 \includegraphics[width=8cm]{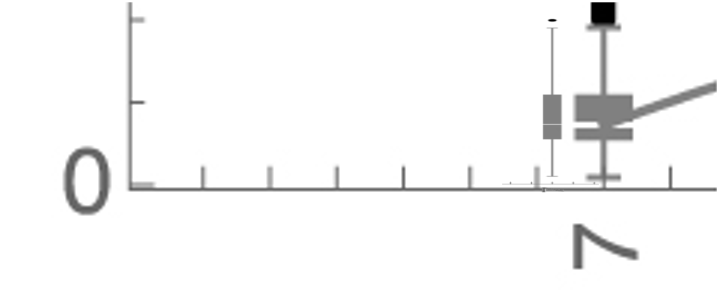}
 \includegraphics[width=8cm]{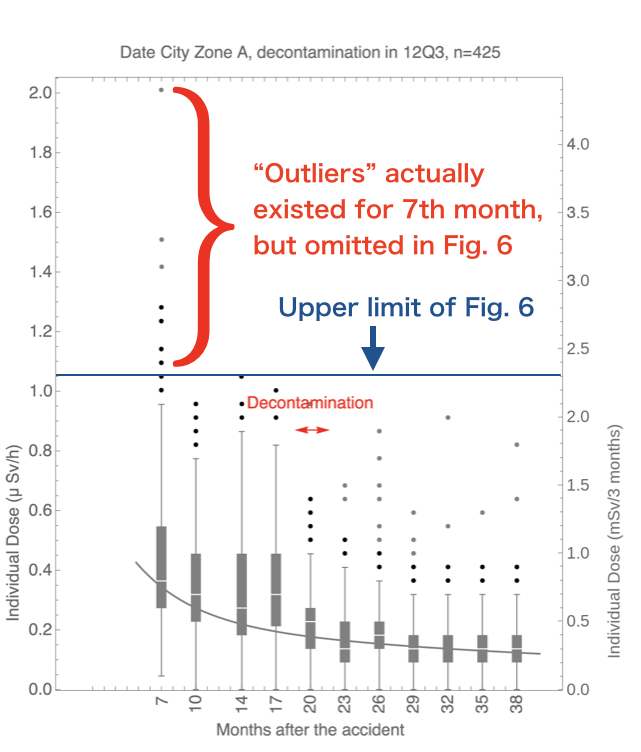}
    \caption{Left: we compare the first box-and-whisker plot of Fig.\ 7 (enlarged) and that of Fig.\! 6 (vertically shrunken and overlaid alongside) of \cite{MH17}. Right: we reconstructed outliers above the first box-and-whisker plot of Fig.\! 6,
    based on the outliers for the same period in Fig.\! 7. The error of the factor 0.46 (see item~(\ref{item.conversion}))
    between the two figures was taken into account.}
  \label{fig:6and7}
\end{figure}
As Fig.\! 7 shows the cumulative doses, the first period contains the doses received during the
 3 months, hence this must coincide (after converting the units) with the first period of Fig.\! 6.
In correspondence to the outliers for the first period of Fig.\! 7,
there must be many outliers above $1.0\,\mu\svh$ for the same period in Fig.\! 6.
A reconstruction of these outliers, taking the error of the factor 0.46 between Figs.\! 6 and 7
(see item~(\ref{item.conversion})) into account, is given in Figure~\ref{fig:6and7} of this Letter.
Evidently, these significant data were truncated from Fig.\! 6, without any comment or explanation.

 \item Section 5 claims that ``\textit{[R]egardless of the magnitude of the ambient dose rates,
 the difference in decontamination method, or whether or not decontamination was carried out,
 the reduction function was the same throughout Date City}''. The authors did not show it.
 Rather, they just assumed the same reduction function in different zones.

 \item The authors' conflicts of interest are not properly declared.
 Date City has asked the authors to carry out the research and write scientific papers, and one of the authors (M.\! Miyazaki)
 is a policy adviser of Date City (see Ethics statement of Ref.~\cite{MH16}).
% \item repeat comments in \cite{Kurokawa18}?
\end{enumerate}

As we have shown, the paper \cite{MH17} contains many serious errors, and/or inappropriate handling of data.
They inevitably affect the main conclusions of the paper, which must be regarded invalid in the light of
the issues pointed out in this Letter.

%The paper, together with its first part \cite{MH16}, went through
%formal investigations by the universities the authors are affiliated to \cite{FMU19July, UT19July},
%but these investigations acknowledged only one technical error in \cite{MH17}
%(the authors have been cleared of both research misconduct and violation of the Ethical Guidelines
%established by the Japanese government \cite{MEXT_misconducts}),
%without mentioning to other issues pointed out in \cite{Kurokawa18}.

\subsubsection*{Acknowledgements}
We thank Ms.\! Akemi Shima for providing us with
the public documents obtained through her Freedom Of Information requests.

Some of the issues pointed out in this Letter were discussed in Refs.~\cite{Kurokawa19,KT19-1,KT19-2}, in the Japanese magazine \textit{KAGAKU}.
This Letter is composed as an original article in English,
with a permission of the publisher of \textit{KAGAKU} (Iwanami Shoten, Publishers).
We also thank the \textit{KAGAKU} Editorial Office for opportunities to discuss this work.

\subsubsection*{Conflicts of interest}
Y.T.'s employment until February 2020 was funded through Programma per giovani ricercatori, anno 2014
``Rita Levi Montalcini'' of the Italian Ministry of Education, University and Research (MIUR).

\end{document}